 \documentclass{emulateapj}
 \usepackage{apjfonts}

\shortauthors{T\"or\"ok \& Kliem}
\shorttitle{Eruptions of kink-unstable flux ropes}

\begin{document}


 \slugcomment{Received 2005 March 4; accepted 2005 July 18}
 \journalinfo{\parbox{\hsize}{\sc\relax 
               ms19378, July 1, 2005}} 

\title{Confined and ejective eruptions of kink-unstable flux ropes} 

\author{T. T\"or\"ok  \altaffilmark{1} and 
        B. Kliem      \altaffilmark{2}}

\altaffiltext{1}{Mullard Space Science Laboratory, University College
                 London, Holmbury St. Mary, Dorking, Surrey RH5 6NT, UK; 
 		 \texttt{tt@mssl.ucl.ac.uk}} 
\altaffiltext{2}{Astrophysical Institute Potsdam, 
                 An der Sternwarte 16, 14482~Potsdam, Germany; 
                 \texttt{bkliem@aip.de}} 

\begin{abstract}

The ideal helical kink instability of a force-free coronal magnetic flux
rope, anchored in the photosphere, is studied as a model for solar
eruptions. Using the flux rope model of \cite{Tit:Dem-99} as the initial
condition in MHD simulations, both the development of helical shape and
the rise profile of a confined (or failed) filament eruption (on 2002 May
27) are reproduced in very good agreement with the observations. By
modifying the model such that the magnetic field decreases more rapidly
with height above the flux rope, a full (or ejective) eruption of the
rope is obtained in very good agreement with the developing helical
shape and the exponential-to-linear rise profile of a fast coronal mass
ejection (CME) (on 2001 May 15). This confirms that the helical kink
instability of a twisted magnetic flux rope can be the mechanism of the
initiation and the initial driver of solar eruptions. The agreement of
the simulations with properties that are characteristic of many eruptions
suggests that they are often triggered by the kink instability. The
decrease of the overlying field with height is a main factor in deciding
whether the instability leads to a confined event or to a CME.

\end{abstract}


\section{Introduction}
\label{intro}

Large-scale solar eruptions occur as flares, filament (or prominence)
eruptions, and coronal mass ejections (CMEs). Despite their different
observational appearance, it is believed that these phenomena are
manifestations of the same physical processes, which involve the
disruption of the coronal magnetic field. Indeed, in the largest
eruptions (eruptive flares) usually all three phenomena are observed. The
theory of the main phase of such events, referred to as the ``standard
model'' of eruptive flares \cite[e.g.,][]{Shibata-99}, is quite well
established. However, their initiation as well as the mechanism of upward
acceleration are still unclear. A variety of theoretical models have been
proposed to explain the impulsive onset and initial evolution of solar
eruptions \cite[see, e.g.,][]{Forbes-00}.

Here we focus on a flux rope instability model. This is motivated by the
observation that solar eruptions often show the phenomenology of a
loop-shaped magnetic flux system with fixed footpoints at the coronal
base and signatures of magnetic twist. Furthermore, erupting filaments
very often develop a clearly helical axis shape in the course of the
eruption, which is the characteristic property of the helical kink
instability of a twisted magnetic flux rope. The instability occurs if
the twist, a measure of the winding of the field lines about the flux
rope axis, exceeds a critical value \citep{Hoo:Pri-81}.

In coronal applications, the simplifying assumption of straight,
cylindrically symmetric flux ropes has nearly always been used so far.
Only very recently, \citeauthor*{Toer:al-04} (\citeyear{Toer:al-04},
hereafter Paper~I) performed the first detailed study
of the kink instability of an arched flux rope, line-tied to the
photosphere, using the analytical model of a force-free coronal flux rope
developed by \citeauthor{Tit:Dem-99} (\citeyear{Tit:Dem-99}, hereafter
TD) as the initial condition in 3D ideal MHD simulations. They have shown
that this model relaxes to a numerical equilibrium very close to the
analytical expressions in the case of subcritical twist and that the
helical kink instability develops for supercritical twist
\cite[see also][]{Fan:Gib-03,Fan:Gib-04}.

In the course of the instability, a helical current sheet, wrapped around
the kinking and rising flux rope where it pushes into the surrounding
field, and a vertical current sheet below the rope (which has no
counterpart in the cylindrically symmetric case) are formed. A vertical
current sheet below rising unstable magnetic flux is the central element
in the standard model of eruptive solar flares. Further essential
features of solar eruptions could be reproduced in the simulations, as
for example the formation of transient soft X-ray sigmoids
\cite*[][Paper~II]{Klie:al-04}. However, a full eruption of the
configuration has not yet been obtained; the flux rope reached an
elevation of only about twice its initial height.

Here we present further developments of these simulations to
substantiate our suggestion in Papers~I and II that
the kink instability of a coronal magnetic flux rope is a possible
trigger mechanism of solar eruptions. The instability was first
suggested as the trigger of (confined and ejective) prominence
eruptions by \cite{Sakurai-76}, but has recently been
generally regarded as a possible explanation only for confined
events \cite[e.g.,][]{Ger:Hoo-03}. The new simulations show that the
instability can also trigger full eruptions.

\section{Numerical model}
\label{numerics}

\begin{figure}[t]                                                 
 \centering
 \includegraphics[width=3.25in]{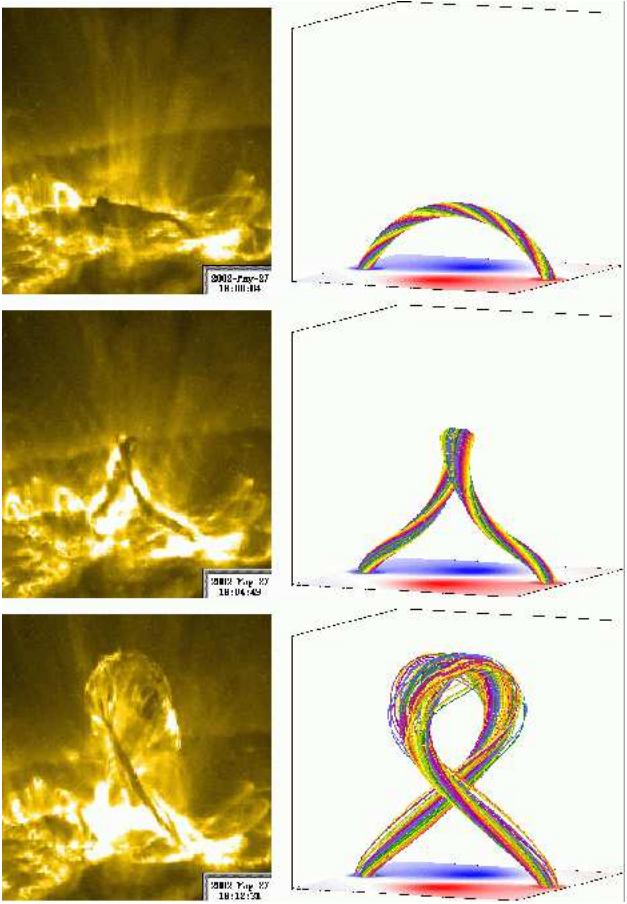} 
\caption[]
{\rule{0pt}{\baselineskip}
 \emph{Left:} \textsl{TRACE} 195~{\AA} images of the confined filament
              eruption on 2002 May 27.
 \emph{Right:} magnetic field lines outlining the core of the
              kink-unstable flux rope (with start points in the bottom
	      plane at circles of radius $b/3$) at $t=0$, 24, and 37.
	      The central part of the box (a volume of size $4^3$) is
	      shown, and the magnetogram, $B_z(x,y,0,t)$, is included.
 \par\vspace{.5\baselineskip}
 }
\label{fig:fila2_loop}
\end{figure}

\placefigure{fig:fila2_loop}

We integrate the compressible ideal MHD equations using the simplifying
assumptions of vanishing plasma-beta, $\beta=0$,
%
%
%
and vanishing gravity,
which are identical to Eqs.~(2--5) in Paper~I.
Setting $\beta=0$ is usually a very good approximation in the lower
and middle corona of active regions, the source re\-gion of most eruptions,
where estimates yield $\beta\!\sim\!10^{-3}\dots10^{-2}$. Both the pressure
gradient force and the gravity force influence the rise characteristics
of the unstable flux rope and the energy partition in the development of
the instability. However, the basic characteristics of the instability
are well described by
%
%
the equations used whenever the
Lorentz force dominates, as is the case in the initial phase of solar
eruptions.
Magnetic reconnection occurs in the simulations due to numerical diffusion
if current sheets steepen sufficiently.

As in Papers~I and II, we use the approximate analytical force-free
equilibrium of an arched, line-tied, and twisted flux rope by TD as the
initial condition for the magnetic field. The flux rope is modelled by
the upper section of a toroidal ring current, partly submerged below the
photosphere, whose Lorentz self-force is balanced by a pair of fictitious
subphotospheric magnetic charges. A fictitious subphotospheric line
current at the toroidal symmetry axis is included to achieve a finite
twist everywhere in the system. See TD for a detailed description of the
model.

The initial density distribution can be freely specified; we choose it
such that the Alfv\'en velocity in the volume surrounding the flux rope
decreases slowly with height:
$\rho_0\propto|\mathbf{B}_0(\mathbf{x})|^{3/2}$ (see
Fig.\,\ref{fig:compare_b} below). The system is at rest at $t=0$,
except for a small upward velocity perturbation, which is localized at
the flux rope apex in a sphere of radius equal to the minor radius, $b$,
of the rope. Lengths, velocities, and times are normalized, respectively,
by the initial flux rope apex height, $h_0$, the initial Alfv\'en
velocity at the apex, $v_\mathrm{A0}$, and the corresponding Alfv\'en
time, $\tau_\mathrm{A}=h_0/v_\mathrm{A0}$.

\begin{figure}[t]                                                 
 \centering
 \includegraphics[width=3.25in]{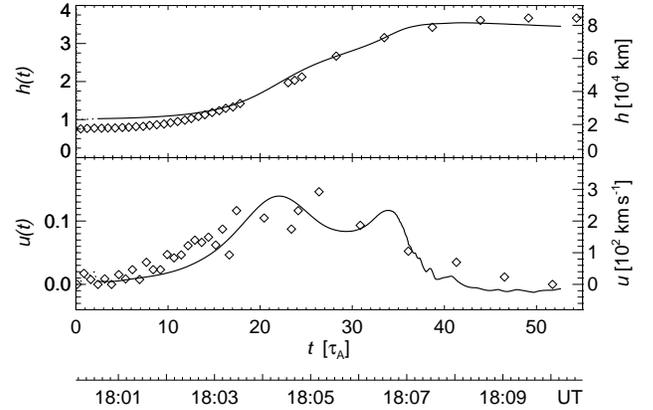}
\caption[]
{Comparison of height, $h(t)$, and velocity, $u(t)$, of the flux rope
 apex in the simulation (solid lines; initial perturbation is dotted) with
 the corresponding values of the filament eruption in
 Fig.~\ref{fig:fila2_loop} (data from Fig.~3 of \citeauthor{Ji:al-03}
 [\citeyear{Ji:al-03}]) overplotted as diamonds (the height data observed
 before 18:04~UT were smoothed here, resulting in reduced velocity scatter).
 See text and Table~\ref{tab:params} for the scaling of the dimensionless
 simulation variables (left axes) to the observed values (right axes).
 \par\vspace{.5\baselineskip}
 }
\label{fig:fila2_apex}
\end{figure}

\placefigure{fig:fila2_apex}

\section{Simulation of a confined eruption}
\label{sec:confined}

The eruption of an active region filament (on 2002 May 27), which was
accompanied by an M2 flare but did not lead to a CME, was described by
\cite{Ji:al-03}. The filament started to rise rapidly and developed a
clear helical shape, as is often observed; however, the ascent was
terminated at a projected height of $\approx$\,80~Mm
(Figs.\,\ref{fig:fila2_loop}, \ref{fig:fila2_apex}). Such confined
filament eruptions are not uncommon \citep{Rust-03}.

In order to model this event, we consider a kink-unstable configuration
which is very similar to the case of an average flux rope twist of
$4.9\pi$ studied in detail in Paper~I. The line current is reduced
by about one third to enable a higher rise, but the average
twist, $\Phi=5.0\pi$ here, is kept to reproduce the helical shape.
(In the TD model, this fixes the minor radius to $b=0.29$, 8~percent
larger than in the reference run in Paper~I.) The sign of
the line current is chosen to be positive to conform to the
apparent positive (right-handed) helicity of the observed filament.
Furthermore, we increase the numerical diffusion and prevent the density
from becoming negative, which permits us to follow the
evolution of the system for a considerably longer time.
Otherwise, the magnetic configuration and the numerical settings are
the same as in Paper~I.

As in our previous simulations, the upwardly directed  kink instability
leads to the ascent and helical deformation of the flux rope as well as
to the formation of current sheets (see Fig.\,3 in Paper~I).
In Fig.\,\ref{fig:fila2_loop} we compare the evolution of the helical
shape of the flux rope with \textsl{Transition Region and Coronal
Explorer (TRACE)} observations of the filament eruption. The evolution is
remarkably similar. Figure\,\ref{fig:fila2_apex} shows that the principal
features of the observed rise are also matched. After an
exponential rise the flux rope comes to a stop at $\approx3.5~h_0$.
A first deceleration occurs as the current density in the helical current
sheet above the apex begins to exceed the current density in the flux
rope ($t>22$). The subsequent upward push ($t>30$) results from the
reconnection outflow in the vertical current sheet below the rope.
Finally, the rise is terminated by the onset of magnetic reconnection in
the current sheet above the rope, which progressively cuts the rope field
lines ($t\ga33$). The reconnection outflows expand the top part of the
rope in lateral direction, as seen both in observation and simulation.

\placetable{tab:params}
\begin{deluxetable}{ccccccccc}                                     
 \tablecaption{Parameter Settings
               \label{tab:params}}
 \tablecolumns{9}
 \tablewidth{0pt}
 \tablehead{
  \colhead{} &
   \multicolumn{4}{c}{Simulation Parameters} &
    \multicolumn{3}{c}{Scaling Parameters} &
     \colhead{} \\
  \colhead{$\!$Sect.$\!$} &
   \colhead{$\!\Phi/\pi$} &
    \colhead{$b$} &
     \colhead{$\!\!\eta(2)\!\!$} &
      \colhead{$L$} &
        \colhead{$h_0$} &
         \colhead{$\tau_A$} &
          \colhead {$|\mathbf{B}_0(h_0)|$}&
           \colhead{$\!\!W$}\\
  \colhead{} & \colhead{} & \colhead{} & \colhead{} & \colhead{} & 
  \colhead{(Mm)} & \colhead{(s)} & \colhead{(G)} & \colhead{$\!\!$(erg)}
  }
 \startdata
  3 & $\!~5.0$ & 0.29 & 0.83 & 10 & ~23 & 11.5 & 200  & $\!\!10^{31}$ \\
  4 & $\!-5.0$ & 0.33 & 1.54 & 32 & 115 & 111  & 10-40& $\!\!10^{31\mbox{--}32}$
 \enddata
 \tablecomments{The expression
  $\eta(z)=-z\,d\ln{B_\mathrm{ex}(0,0,z,0)}/dz$ is the `decay index' of
  the `external' field (excluding the contribution by the ring current),
  $L$ is the box size, and $W$ is the released magnetic energy.
  The runs are equal in grid resolution in the central part of the
  box, $\Delta=0.02$, major rope radius, $R=1.83$, and distance of
  the fictitious magnetic charges from the $z$ axis, $l=0.83$.
  }
\end{deluxetable}

Using the scaling to dimensional values given in Table~\ref{tab:params},
good quantitative agreement with the rise profile is obtained
(Fig.~\ref{fig:fila2_apex}), and the release of magnetic energy in this
run of 5~percent
%
%
corresponds to $10^{31}$~erg, a reasonable value for
a confined M2-class flare.

The agreement between the observations of the event and our simulation
confirms the long-held conjectures that the development of strongly
helical axis shapes in the course of eruptions can be regarded as an
indication of the kink instability of a twisted flux rope and that
the frequently observed helical fine structures in erupting filaments and
prominences outline twisted fields. Furthermore, it shows that flux ropes
with substantial twist can exist or be formed in the solar
corona at the onset of, or prior to, eruptions.

\section{Simulation of an ejective eruption (CME)}
\label{sec:ejective}

\begin{figure}[t]                                                 
 \centering
 \includegraphics[width=3.25in]{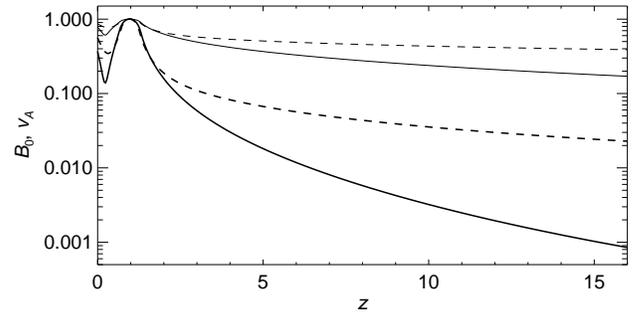}   
\caption[]
{Normalized initial magnetic field strength (\emph{thick lines}) and
 Alfv\'en velocity (\emph{thin lines}) vs.\ height for the configurations
 described in
 Sect.\,\ref{sec:confined} (original TD model; \emph{dashed lines}) and
 Sect.\,\ref{sec:ejective} (modified TD model; \emph{solid lines}).
 \par\vspace{.5\baselineskip}
 }
\label{fig:compare_b}
\end{figure}

The full eruption of the kink-unstable flux rope in the TD model is
prevented by the strong overlying field, which is dominated by the line
current. It is possible to obtain an eruptive behaviour of the flux rope
by removing the line current; however, such a modification leads to an
infinite number of field line turns at the surface of the flux rope
\citep{Rous:al-03}. In order to avoid this problem, we
replaced the line current by a pair of subphotospheric
dipoles \cite[as used in][]{Toer:Kli-03}. The position of the dipoles is
chosen such that the field lines of the dipole pair passing through the
flux rope match the curvature of the rope as closely as possible. The
resulting equilibrium yields a finite twist everywhere in the system, but
the magnetic field overlying the flux rope now decreases significantly
faster with height than in the original TD model
(Fig.\,\ref{fig:compare_b}). By varying the dipole moments or the minor
radius $b$, one can adjust the average twist within the flux rope.

Choosing suitable dipole moments and $b=0.6$, but otherwise the same
parameters of the TD model as in Sect.\,\ref{sec:confined}, we
first checked that the modified configuration relaxes to a nearby stable
equilibrium for subcritical twist, $\Phi=2.7\pi$.
Next a configuration with supercricital twist, $\Phi=-5.0\pi$, is
considered, obtained by changing the minor radius to $b=0.33$ and
reversing the dipole moments. The sign of the helicity corresponds to the
2001 May 15 event discussed below; it has no influence on the rise,
$h(t)$, of the flux rope apex. The numerical parameters of the simulation
are the same as in Sect.\,\ref{sec:confined}, except for a considerably
larger simulation box and a smaller level of numerical diffusion, which
this system permitted.

\begin{figure*}[t]                                                
 \centering
 \includegraphics[width=.9\textwidth]{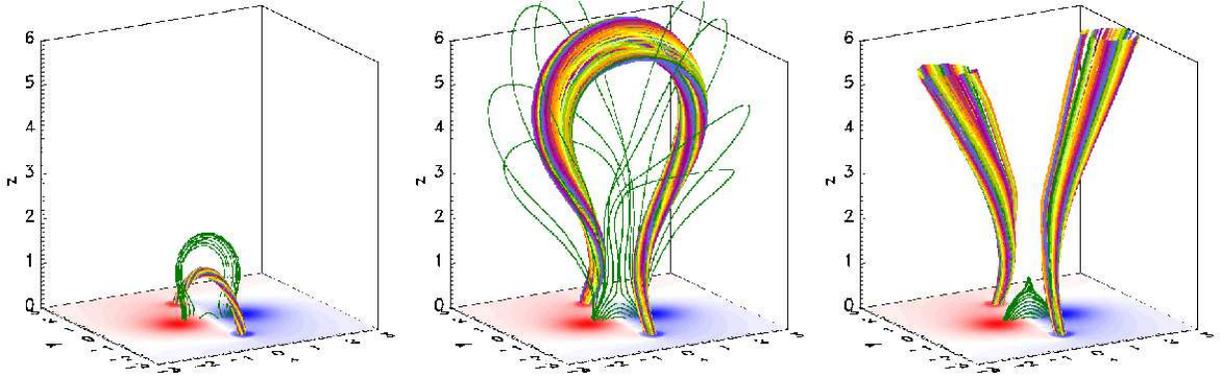} 
\caption[]
{Magnetic field lines of the kink-unstable modified TD model at
 $t=0$ (\emph{left}), $t=30$ (\emph{center}), and $t=43$ (\emph{right}).
 The magnetogram, $B_z(x,y,0,t)$, is included.
 Field lines started at a circle of radius $b/3$ in the bottom plane show
 the core of the flux rope.
 Additional green field lines, also with identical start points in all
 panels, indicate the formation of ``post flare loops'' with a cusp by
 reconnection.
 The hyperbolic point of the field at the $z$-axis (magnetic X-point)
 lies at $z\approx0.2$, 0.6, and 1.1, respectively.
 \par\vspace{.5\baselineskip}
 }
\label{fig:db1_loop}
\end{figure*}

The helical kink instability also develops in the modified model.
However, the flux rope now exhibits a much stronger expansion
(Fig.\,\ref{fig:db1_loop}), which is not slowed down.
The initially exponential rise is followed by a rise with approximately
constant and locally super-Alfv\'enic velocity, until the rope encounters
the top of the simulation box (at $t\approx80$).
The helical current sheet remains very weak on top of the flux rope apex
and no significant amount of reconnection occurs here. On the other hand,
the vertical current sheet now steepens in a large height range. Magnetic
reconnection commences in this sheet at the beginning of the exponential
phase and rises in tandem with the ascent of the flux rope, particularly
closely during the exponential phase. Since the flux rope expands
continuously during this phase (instead of being compressed by the upward
reconnection outflow below it) and moves away from the forefront
of the outflow region afterwards, the ideal instability of the flux rope
appears to be the driver of the closely coupled processes. Cusp-shaped
field lines are formed throughout the evolution
(Fig.\,\ref{fig:db1_loop}) but most prominently in the late phase, in
agreement with soft X-ray observations of eruptive flares.

\placefigure{fig:compare_b}

The full eruption of the flux rope in the modified TD model must be
enabled by the weaker overlying field, since all other parameters are
identical, or very close, to
%
%
Sect.\,\ref{sec:confined}.

\placefigure{fig:db1_loop}

\begin{figure}[t]                                                
 \centering
 \includegraphics[width=3.25in]{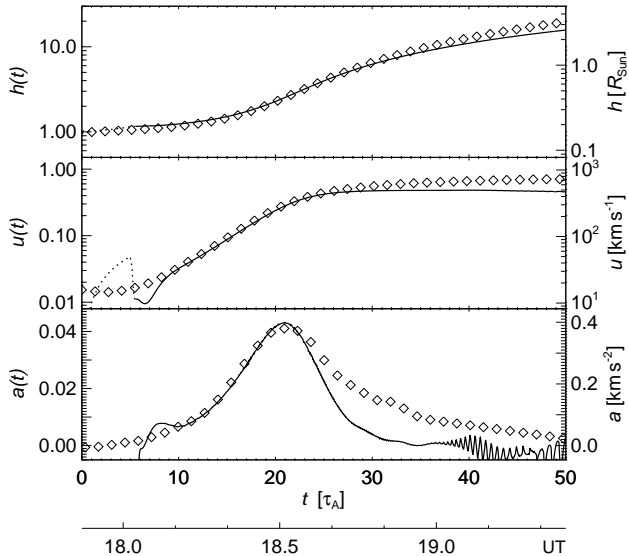}   
\caption[]
{Comparison of the simulation in Sect.~\ref{sec:ejective} with the CME on
 2001 May 15 in the same format as in Fig.~\ref{fig:fila2_apex},
 including the acceleration $a(t)$. Diamonds show the apex motion of the
 erupting prominence/the CME core (data are from Fig.\,6a,\,c of
 \citeauthor{Mari:al-04} [\citeyear{Mari:al-04}]).
 \par\vspace{.5\baselineskip}
 }
\label{fig:db1_apex}
\end{figure}

\placefigure{fig:db1_apex}

In Fig.\,\ref{fig:db1_apex} we compare the rise of the flux rope apex in
the simulation with the rise of the apex of a well observed eruptive
prominence on 2001 May 15, which occurred in a spotless region slightly
behind the limb and led to a fast CME (peak velocity of leading edge
$\approx$\,1200~km\,s$^{-1}$) and a long-duration flare (X-ray class C4).
As described by \cite{Mari:al-04}, the eruptive prominence
developed a helical shape, analogous to the middle panels in
Fig.\,\ref{fig:fila2_loop} with reversed handedness, and it showed the
typical rise characteristics of a fast CME
\cite[initially exponential or exponential-like rise, followed by
approximately linear rise;][]{Vrsnak-01}. For
this comparison, we first scaled the Alfv\'en time such that the duration
of the exponential rise phase is matched, $\tau_\mathrm{A}=111$~s, and
shifted the time axis accordingly. Then we scaled the length unit such
that the apex height at the point of peak acceleration in the simulation
($t=21$) equals the height of the prominence at the resulting observation
time (see bottom axis), i.e., $h_0=115$~Mm. This fixes the scaling of the
velocity and acceleration amplitudes. Apart from a somewhat more gradual
decrease of the observed acceleration after the peak, excellent
qualitative and quantitative agreement is obtained, demonstrating (as in
Sect.~\ref{sec:confined}) that the kink instability yields the growth
rate required by the observed rise profile for the twist indicated by the
observed helical shape. The slight difference in the late acceleration
profile may have many origins, for example, a different height profile of
the field strength, or a simultaneous expansion of the overlying field
enforced by photospheric flows in the observed event.

The simulation shows a strong magnetic energy release,
25~percent of the initial value,
%
%
which agrees with
the magnitude observed in ejective solar eruptions
\citep{Forbes-00,Emsl:al-04}. The considered event may have released
magnetic energy of order $\sim(10^{31}\mbox{--}10^{32})$~erg (the low
X-ray class resulted from footpoint occultation). Using the above scaling
for $h_0$, this energy release is reproduced for
$B_0\sim(10\mbox{--}40)$~G, consistent with expected averages of the
coronal field strength over the large length scales involved.

The simulation could also be scaled to a filament eruption that was
associated with an X-class flare and a very fast CME
\cite[on 2004 November 10; see][]{Will:al-05}. 

A line-tied flux rope was found to erupt in a few previous simulations
\citep{Amar:al-00,Amar:al-03a,Amar:al-03b}. The present simulations,
through their agreement with characteristic properties of solar
eruptions, \emph{identify a mechanism} for the process, confirming the
original suggestion by \cite{Sakurai-76}. They also demonstrate the
importance of the height dependence of the overlying field  for the
evolution of the instability into a CME, while \cite{Amar:al-03b} found
the amount of magnetic helicity to be essential. Since
\citeauthor{Amar:al-03b} built up the helicity by rotating the main
photospheric polarities, which simultaneously expands the overlying field
\citep{Toer:Kli-03}, both results are fully consistent with each other.

\section{Conclusions}
\label{concl}

Our MHD simulations of the kink instability of a coronal magnetic flux
rope reproduce essential properties---an initially exponential rise with
the rapid development of a helical shape---of two well observed solar
eruptions, one of them confined, the other ejective. The subsequent
approximately linear rise of the ejective eruption is reproduced as well.
Since these features are characteristic properties of many solar
eruptions \citep{Vrsnak-01}, we regard the kink instability of coronal
magnetic flux ropes as the initiation mechanism and initial driver of
many such events. A sufficiently steep decrease of the magnetic field
with height above the  flux rope permits the process to evolve into a
CME.

\acknowledgements

We acknowledge constructive comments by the referee
and thank H.~Ji and B.~Vr\v{s}nak for the observation data in
Figs.~\ref{fig:fila2_apex} and \ref{fig:db1_apex}, respectively.
This work was supported by grants 50\,OC\,9706 (DLR), MA~1376/16-2 (DFG),
and HPRN-CT-2000-00153 (EU).
%
The John von Neumann Institute for Computing, J\"ulich granted computer
time.




\begin{thebibliography}{}
\bibitem[Amari et al.(2003a)]{Amar:al-03a}
         Amari, T., et al. 2003a, \apj, 585, 1073
\bibitem[Amari et al.(2003b)]{Amar:al-03b}
         Amari, T., et al. 2003b, \apj, 595, 1231
\bibitem[Amari et al.(2000)]{Amar:al-00}
         Amari, T., Luciani, J. F., Mikic, Z., \& Linker, J. 2000, \apjl,
	 529, L49
 \bibitem[Emslie et al.(2004)]{Emsl:al-04}
          Emslie, A. G., et al. 2004, \jgr, 109, A10104
\bibitem[Fan \& Gibson(2003)]{Fan:Gib-03}
         Fan, Y., \& Gibson, S. E.\ 2003, \apjl, 589, L105
\bibitem[Fan \& Gibson(2004)]{Fan:Gib-04}
         Fan, Y., \& Gibson, S. E.\ 2004, \apj, 609, 1123
\bibitem[Forbes(2000)]{Forbes-00}
         Forbes, T. G. 2000, \jgr, 105, 23\,153
\bibitem[Gerrard \& Hood(2003)]{Ger:Hoo-03}
         Gerrard, C. L., \& Hood, A. W. 2003, \solphys, 214, 151
\bibitem[Hood \& Priest(1981)]{Hoo:Pri-81}
         Hood, A. W., \& Priest, E. R. 1981, Geophys. Astrophys. Fluid
         Dyn., 17, 297
\bibitem[Ji et al.(2003)]{Ji:al-03}
         Ji, H., et al. 2003, \apjl, 595, L135
\bibitem[Kliem et al.(2004)Kliem, Titov, \& T\"or\"ok]{Klie:al-04}
         Kliem, B., Titov, V. S., \& T\"or\"ok, T. 2004, \aap, 413, L23
	 (Paper~II)
\bibitem[Mari{\v c}i\'c et al.(2004)]{Mari:al-04}
         Mari{\v c}i\'c, D., Vr{\v s}nak, B., Stanger, A. L., \& Veronig,
	 A.\ 2004, \solphys, 225, 337
\bibitem[Roussev et al.(2003)]{Rous:al-03}
         Roussev, I. I., et al. 2003, \apjl, 588, L45
\bibitem[Rust(2003)]{Rust-03}
         Rust, D. M. 2003, Adv. Space Res., 32, 1895
\bibitem[Sakurai(1976)]{Sakurai-76}
         Sakurai, T.\ 1976, \pasj, 28, 177
\bibitem[Shibata(1999)]{Shibata-99}
         Shibata, K. 1999, Ap{\&}SS, 264, 129
\bibitem[Titov \& D\'emoulin(1999)]{Tit:Dem-99}
         Titov, V. S., \& D\'emoulin, P. 1999, \aap, 351, 707 (TD)
\bibitem[T\"or\"ok \& Kliem(2003)]{Toer:Kli-03}
         T\"or\"ok, T., \& Kliem, B. 2003, \aap, 406, 1043
\bibitem[T\"or\"ok et al.(2004)T\"or\"ok, Kliem, \& Titov]{Toer:al-04}
         T\"or\"ok, T., Kliem, B., \& Titov V. S. 2004, \aap, 413, L27
	 (Paper~I)
\bibitem[Vr{\v s}nak(2001)]{Vrsnak-01}
         Vr{\v s}nak, B.\ 2001, \jgr, 106, 25\,249
\bibitem[Williams et al.(2005)]{Will:al-05}
        Williams, D. R., T\"or\"ok, T., D\'emoulin, P.,
        van Driel-Gesztelyi, L., \& Kliem, B. 2005, \apjl, 628, L163
\end{thebibliography}
\end{document}